\documentclass[12pt,letterpaper]{JHEP3}
\pdfoutput=1
\usepackage{cite}
\usepackage{epsfig}
\usepackage{graphicx}
\usepackage{subfigure}

\graphicspath{{Figures/}}

\title{Recovering the Negative Mode for \\ Type B Coleman-de Luccia Instantons}

\author{I-Sheng Yang\footnote{isheng.yang@gmail.com}\\
ISCAP and Physics Department \\
Columbia University, New York, NY, 10027 , U.S.A. \\
AND \\
IOP and GRAPPA, Universiteit van Amsterdam, \\
Science Park 904, 1090 GL Amsterdam, Netherlands
}

\abstract{The usual (type A) thin-wall Coleman-de Luccia instanton is made by a bigger-than-half sphere of the false vacuum and a smaller-than-half sphere of the true vacuum.  It has a the standard $O(4)$ symmetric negative mode associated with changing the size of false vacuum region.  On the other hand, the type B instanton, made by two smaller-than-half spheres, was believed to have lost this negative mode.  We argue that such belief is misguided due to an over-restriction on Euclidean path integral.  We introduce the idea of a ``purely geometric junction'' to visualize why such restriction could be removed, and then explicitly construct this negative mode.  We also show that type B and type A instantons have the same thermal interpretation for mediating tunnelings.}

\begin{document}

\section{Introduction}

\begin{figure}
\begin{center}
\includegraphics[width=8cm]{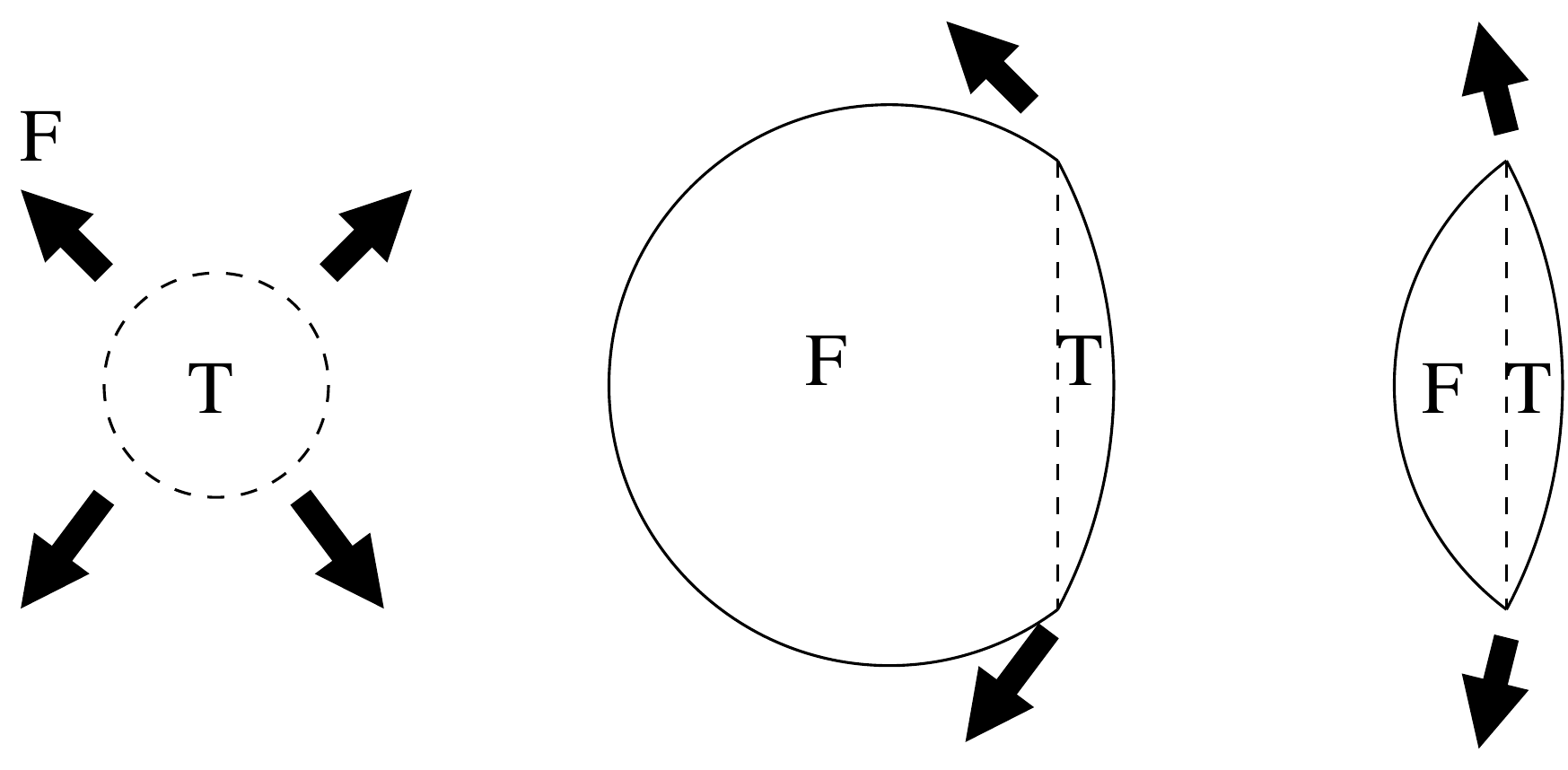}
\caption{From left to right we show the flat space instanton solution with a critical bubble, the type A instanton solution between two de~Sitter vacua, and the type B instanton solution between the same pair of de~Sitter vacua.  The dashed line is the domain wall.  Note that the domain wall tension must be different between the type A and type B instantons if they are between the same pair of de~Sitter spaces.  The arrows represent the deformation that changes the bubble size, which corresponds to the negative mode in the first two cases, but a positive mode for the last case.
\label{fig-CDL}}
\end{center}
\end{figure}

Coleman and de Luccia wrote down the instanton solution that became the paradigm of first order phase transitions with gravity\cite{CDL}.  When the critical bubble is much smaller than the de~Sitter radius of the parent vacuum, the CDL instanton has almost the entire 4-sphere of the false vacuum, and a small bubble of the true vacuum.  It is very similar to the Coleman instanton in flat space\cite{Col77}.  In the thin-wall approximation, the instanton geometry contains a kink at the domain wall between the true and false vacua, as shown in Fig.\ref{fig-CDL}.  In the conventional analysis, there is a negative mode corresponding to moving the domain wall (together with the kink), which is similar to changing the bubble size in the flatspace version.

Interestingly, a continuous parameter change from the above ``type A'' instanton leads to the ``type B'' instanton.  As depicted in Fig.\ref{fig-CDL}, the type B instanton contains two smaller-than-half spheres.  Such solution is troublesome in two aspects.  First, it appears to lose the usual negative mode corresponding to the change of bubble size\cite{MarTur07}.  Second, having less than half of the false vacuum de~Sitter 4-sphere makes it difficult to interpret the phase transition as nucleating a bubble.

The existence and uniqueness of the negative mode is a criterion for the instantons to mediate vacuum transitions.  For the Coleman instanton without gravity, it was proved in\cite{Col87}.  Considering the gauge theory nature of gravity, the CDL instanton needs certain appropriate mode reduction process, otherwise there will be spurious modes\cite{TanSas92,Tan99,Lav99}.  A tentative existence plus uniqueness proof was presented in\cite{KhvLav00}\footnote{It is tentative because a not-fully-justified analytical continuation for the uniqueness proof, and a possible sign problem of $Q_E$ in the existence proof as pointed out in\cite{BatLav12}.}. Using similar techniques, numerical thick-wall examples of both type A and type B instantons are shown to have exactly one negative mode\cite{GraTur00,Lav06,BatLav12}.  Unfortunately, such framework provides only the existence but cannot clearly demonstrate which physical deformation the negative mode corresponds to.  No one has explicitly constructed the physical deformation of the negative mode for type B instantons.  Thus the sharp contrast between type A and type B instantons is not fully resolved.

In this paper, we will stick to the thin-wall approximation and introduce the idea of a ``purely geometric junction''.  We explain why this feature is allowed in off-shell configurations of Euclidean path integral.  Employing this feature, we can explicitly construct the physical deformation corresponding to the negative mode for type B instantons.  Just like for the type A instantons, the deformation is still the change of bubble size.\footnote{It is actually the real change of bubble size---it increases/decreases the volume of the true/false vacuum region and vice versa.  The mode shown in Fig.\ref{fig-CDL} does that for a type A instanton, but for a type B it increases/decreases the volume for both vacuum regions at the same time.}  In light of this, we see no reason to treat them differently.  In fact, we will show that in the thermal interpretation\cite{BroWei07}, it represents the phase transition similarly to how a type A instanton represents the reverse transition\cite{LeeWei87}. 

The structure of this paper goes like the following.  In Section \ref{sec-CDL}, we review the basic solution of a thin-wall CDL instanton and the missing negative mode for type B.  In Section \ref{sec-Nmode} we justify the usage of purely geometric junctions and explicitly construct this negative mode.  In Section \ref{sec-interpretation} we discuss how to interpret the type B instanton as mediating phase transitions.  Finally we summarize and conclude in Section \ref{sec-conclusion}.  In Appendix \ref{sec-app} we provide the simplest thick wall construction to address possible concerns and further justify our usage of purely geometric junctions.

\section{Two Types of CDL Instantons}
\label{sec-CDL}

Consider a scalar field with the following potential,
\begin{equation}
V(\phi) = \lambda^2(\phi^2-\phi_0^2)^2 - \frac{\Delta V\phi}{2\phi_0}~.
\label{eq-pot}
\end{equation}
We have a true vacuum and a false vacuum at
\begin{equation}
\phi_T \approx \phi_0~, \ \ \ \  \phi_F \approx -\phi_0~,
\end{equation}
with energy difference roughly $\Delta V$, and a domain wall separating them with tension given by
\begin{equation}
\sigma \approx \int_{\phi_F}^{\phi_T} \sqrt{2V} d\phi~.
\end{equation}

Given that $\Delta V\ll\lambda^2\phi_0^4$, a false vacuum background can nucleate a thin-wall bubble of true vacuum, which will then expand and complete the phase transition.  The rate of this nucleation process is given by (keeping only the exponent)
\begin{eqnarray}
\Gamma_{F\rightarrow T} &=& e^{S_{I}-S_{F}}~, \\
S &=& \int L_m~dx^4 = \int 
\left(\frac{1}{2}(\partial\phi)^2+V(\phi)\right) dx^4~.
\end{eqnarray}
$S$ stands for the Euclidean action: $S_F$ for the background false vacuum and $S_I$ for the instanton solution that contains a bubble of true vacuum.  The 4D Euclidean configuration of the instanton solution is a 3-sphere of domain wall, filled with the true vacuum and surrounded by the false vacuum. One can easily write down the action difference,
\begin{equation}
(S_{I}-S_{F}) = 2\pi^2 r^3 \sigma - \frac{\pi^2}{2}r^4\Delta V~.
\end{equation}
It has a maximum at
\begin{equation}
r_c = 3\frac{\sigma}{\Delta V}~.
\label{eq-rc}
\end{equation}
Varying $r$ is an unique negative mode that signifies this configuration being the leading saddle point contribution that mediates the phase transition.

Including gravity, the Euclidean action becomes:
\begin{equation}
S = \int \sqrt{g} dx^4 \left(L_m - \frac{M_p^2}{2}\mathcal{R}\right)~,
\end{equation}
where the matter action is
\begin{equation}
L_m = \frac{1}{2}g_{\mu\nu}\partial^\mu\phi\partial^\nu\phi+V~.
\end{equation}
For simplicity, in this paper we will focus on the scenario that both vacua has positive energy by adding a constant term to $V$, such that
\begin{equation}
V(\phi_T) = \frac{3M_p^2}{R_T^2}~, \ \ \ \  V(\phi_F)=\frac{3M_p^2}{R_F^2}~.
\end{equation}
The Euclidean action of the false vacuum configuration is simply
\begin{equation}
S_F = \left(V(\phi_F)-\frac{6M_p^2}{R_F^2}\right)~
\mathcal{V}(R_F,{\rm full})~.
\end{equation}
The two terms in the bracket are the field and gravity contributions, and the last factor stands for the 4-volume of a 4-sphere with radius $R_F$. 

The instanton is the matching of two 4-spheres with radii $R_T$ and $R_F$ at some junction radius $r$, where the domain wall resides.  Its action is given by
\begin{equation}
S_I = 2\pi^2 r^3 \sigma + L_{g,wall}(R_F,R_T,r) +
\left(V(\phi_F)-\frac{6M_p^2}{R_F^2}\right)~\mathcal{V}(R_F,r)+
\left(V(\phi_T)-\frac{6M_p^2}{R_T^2}\right)~\mathcal{V}(R_T,r)~.
\end{equation}
The last two terms are the combined contribution of field and gravity from two ``shells'' of the true and false vacua.  The first term is the field contribution at the junction, namely the domain wall tension times the wall area.  The second term is the gravitational contribution from the junction\footnote{Note that this term comes from Eq.~(\ref{eq-Lgwint}), in which the Racci scalar $\mathcal{R}$ receives a delta function contribution from the second derivative term since the first derivative is discontinuous across the junction.  The common practice is to integrate by part and turn this term into a Gibbons-Hawking (surface) term and a modification to the Einstein-Hilbert (volume) term.  We will refrain from doing that so the physical meaning of each term remains clear.},
\begin{equation}
L_{g,wall}(R_F,R_T,r) = 6\pi^2M_p^2r^2
\left(\pm\sqrt{1-\frac{r^2}{R_F^2}}\pm\sqrt{1-\frac{r^2}{R_T^2}}\right)~.
\label{eq-Lgw}
\end{equation}
It takes the plus sign when the corresponding side is a portion of 4-sphere that contains a full equator 3-sphere, namely the bigger portion.  Otherwise it takes the minus sign when the corresponding side is a smaller portion.  The same ambiguity appears in $\mathcal{V}(R_F,r)$, which means the volume of a partial 4-sphere bounded by a 3-sphere of radius $r$ and can be either the bigger or the smaller side.  Fortunately as shown in\cite{CDL,Par83} that all these ambiguous terms can be combined to show that without ambiguity, the instanton action is extremized at
\begin{equation}
r_e = \frac{r_c}{\sqrt{1+(R_F^{-2}+R_T^{-2})\frac{r_c^2}{2}
+(R_F^{-2}-R_T^{-2})^2\frac{r_c^4}{16}}}~,
\label{eq-re}
\end{equation}
where $r_c$ is the critical bubble size in flat space, given by Eq.~(\ref{eq-rc}).  When $r_c\ll R_F$, the instanton contains a small portion of the true vacuum and a large portion of the false vacuum, $r_e\sim r_c$, and varying $r$ is a negative mode just as in the case without gravity.  This is the more standard case and called the type A instanton.

An interesting behavior arises when we tune the potential to increase $r_c$, for example by reducing $\Delta V$.  Eq.~(\ref{eq-re}) shows that $r_e$ will eventually become inversely proportional to $r_c$ instead.  At the same time the instanton becomes two smaller-than-half portions of spheres.  This is what we call the type B instanton.  These two cases are drawn in Fig.\ref{fig-CDL}.\footnote{More technically, type A means the false vacuum region is bigger than a half sphere, while type B means the false vacuum region is smaller than a half sphere.  The exact boundary between the two cases is $r_e=R_F$.}  At the level of solving the equation of motion, namely finding the critical point of the Euclidean action, these is no dramatic change between the two cases and Eq.~(\ref{eq-re}) is always valid.  However when one varies the action around this critical point by changing $r$, it corresponds to a negative mode for type A but a positive mode for type B\cite{MarTur07}.

The type B instanton still has an action bigger than $S_F$ (and $S_T$), so people tend to believe that it has at least one negative mode\footnote{Of course, it can also be a local minimum isolated from both vacuum configurations.  That is typically considered less likely.}.  Since the above analysis is restricted to $O(4)$ symmetry and thin-wall approximation, the common intuition is to go beyond either or both of them.  Maybe the disappearance of the radial negative mode signifies the emergence of many more subtle negative modes, and condensing them leads to a thick-wall or less symmetric solution that has only one negative mode.

However, these suggestions are all motivated by the apparent ``disappearance'' of the $O(4)$ symmetric negative mode.  We will explicitly show that such a mode actually still exists.

\section{Recovering the Negative Mode}
\label{sec-Nmode}

Let us reconsider what happened in the extremizing process described in the previous section.  At the extremum, we have
\begin{equation}
2\pi^2 r_e^3 \sigma = -\frac{3}{2} L_{g,wall}(R_F,R_T,r_e)~,
\label{eq-junc}
\end{equation}
which is the Israel junction condition\cite{Isr66}, namely the integrated Einstein equations across a co-dimension one delta function.  It tells us how the tension of the domain wall determines the angle of the geometric kink.

While looking for the negative mode, one varies the position of the domain wall and the geometric kink together to values other than $r_e$.  Those will be off-shell configurations in the path integral, and Eq.~(\ref{eq-junc}) will not hold.  This is totally fine, since the full equations of motions are
\begin{eqnarray}
\phi''+3\frac{\rho'}{\rho}\phi' &=& \frac{\partial V}{\partial \phi}~, 
\label{eq-field} \\
\rho'^2 &=& 1 + \frac{\rho^2}{3M_p^2}
\left(\frac{\phi'^2}{2}-V\right)~, \label{eq-constraint} \\
\rho'' &=& -\frac{2}{3M_p^2}(\phi'^2+V)~, \label{eq-gravity}
\end{eqnarray}
with the metric
\begin{equation}
ds^2 = d\xi^2 + \rho^2d\Omega_3^2~.
\end{equation}
Only Eq.~(\ref{eq-constraint}), the constraint equation, should hold for off-shell configurations.  Eq.~(\ref{eq-junc}) comes from the delta-function integral
\begin{equation}
L_{g,wall} = -\frac{M_p^2}{2} \int_{\bar\xi-\epsilon}^{\bar\xi+\epsilon} 
\mathcal{R} \rho^3 d\xi~,
\label{eq-Lgwint}
\end{equation}
with $\epsilon\rightarrow0$ and $\rho(\bar\xi)=r$.  Its value only involves the $\rho''$ term in $\mathcal{R}$.  So not solving the junction condition just means not solving Eq.~(\ref{eq-gravity}) but still obeys the constraint Eq.~(\ref{eq-constraint}).

What we will do next is qualitatively the same as the usual CDL variation described above.  {\bf In the variation of CDL radial mode, one goes through configurations where the gravity contribution to the junction, Eq.~(\ref{eq-Lgw}), does not match the matter contribution from the domain wall tension.}  For the same token, we shall be allowed to do the following. {\bf Put a geometric junction where there is no domain wall.  It only contributes gravitationally, as if there is a zero tension domain wall.}  The constraint equation demands only that in the true (false) vacuum region, the geometry has to be a portion of the 4-sphere with $R_T$ ($R_F$).\footnote{Strictly speaking this is only true in exact thin-wall situations.  In Appendix \ref{sec-thick} we will provide the thick wall justification.}

\begin{figure}[h]
\begin{center}
\includegraphics[width=8cm]{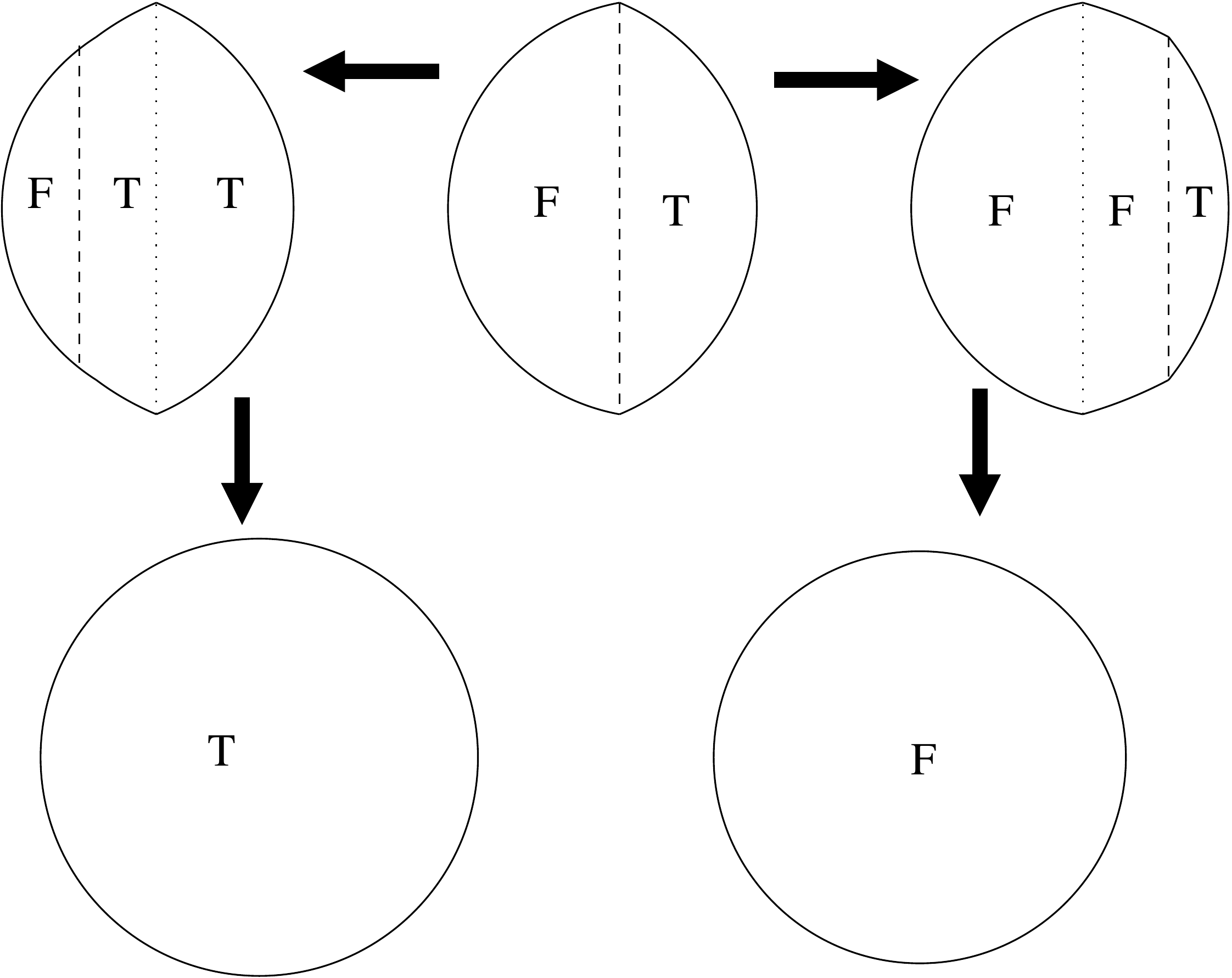}
\caption{The top middle figure is a type B CDL instanton.  We can deform the domain wall (dashed line) away from its critical position to either left or right while leaving a purely geometric junction (dotted line) behind.  This is a negative mode as the action decreases in both directions.   We can further shrink the domain wall and smooth out the junction to recover the true or false vacuum 4-sphere---solutions without negative modes.
\label{fig-deform}}
\end{center}
\end{figure}

With that in mind, let us consider the following solution parametrized by two radii, $r_g$ and $r_w$.  At $r_g$ there is a purely geometric junction, and at $r_w$ we have the usual domain wall with tension $\sigma$ separating the true and false vacuum.  When the purely geometric junction is in the true vacuum region, we have

\begin{eqnarray}
S_I(r_g,r_m) &=& 2\pi^2r_m^3\sigma+L_{g,wall}(R_F,R_T,r_m) + L_{g,wall}(R_T,R_T,r_g) \\
&+& \left(V(\phi_F)-\frac{6M_p^2}{R_F^2}\right)~
\mathcal{V}(R_F,r_m)
\nonumber \\
&+&\left(V(\phi_T)-\frac{6M_p^2}{R_T^2}\right)~
\bigg(\mathcal{V}(R_T,r_m~{\rm to}~r_g) 
+ \mathcal{V}(R_T,r_g)\bigg)~. \nonumber
\end{eqnarray}
When the junction is in the false vacuum region, just switch $T$ and $F$ in the above equation.  All the 4-volume functions $\mathcal{V}$ here are referring to the smaller portion without an equator 3-sphere, which should be obvious from Fig.\ref{fig-deform}.

When $r_g=r_m=r_e$, this is exactly the critical solution of a type B CDL instanton.  Now we can fix $r_g=r_e$ and start to vary $r_m$ to either side, as shown in Fig.\ref{fig-deform}.  We see that the action always decreases\footnote{We skipped the equations or numerical plots here since one can easily observe the following: the change of action is dominated by the domain wall contribution, which is shrinking in both directions of this deformation.}.  We can further shrink $r_m$ to zero to eliminate the true or false vacuum portion, and smooth out the purely geometric junction.   The action is strictly decreasing during the entire process and recovers $S_T$ or $S_F$.

The analysis of radial mode in\cite{MarTur07} was restricted to a single geometric junction that always sticks with the domain wall.  This unnecessary restriction led to a bias that for type B instantons, ``changing the bubble size'' is actually changing the total size of the entire instanton.  As a hindsight, such deformation has no reason to be the relevant negative mode for a tunneling process.  The negative mode should represent two directions that the instanton rolls toward either the true or the false vacuum.  

The physical deformation we show in Fig.\ref{fig-deform} is exactly doing that.  It is ``really'' changing the bubble size---shrinking the false/true vacuum region while expanding the other.  The fact that this deformation corresponds to a negative mode should not be very surprising.

\section{The Thermal Interpretation}
\label{sec-interpretation}

Brown and Weinberg\cite{BroWei07} provided a very accurate picture to interpret how the type A CDL instanton mediates the vacuum transition.  Instead of taking the Euclidean 4-sphere as the global geometry, they described it as a horizon 3-volume times a compact coordinate from the finite temperature.  One side of the equator is the horizon volume before tunneling, and the other side is the same horizon volume after tunneling.

\begin{figure}
\begin{center}
\includegraphics[width=8cm]{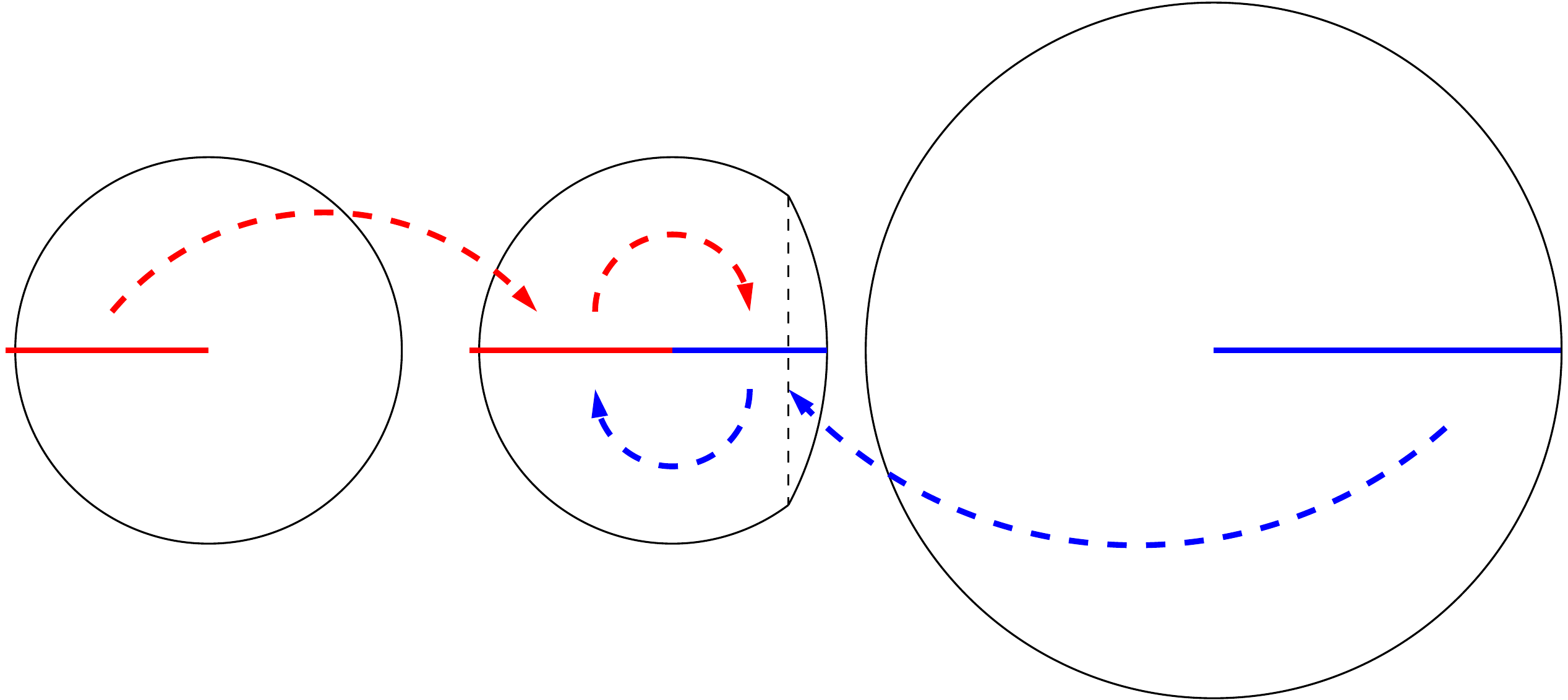}
\caption{The downward (red, left to right) and upward (blue, right to left) vacuum transitions mediated by the type A CDL instanton.  Both processes are shown by two dashed-arch-arrow, one for the thermal fluctuation from the initial horizon volume to the configuration of the corresponding (almost) semi-3-sphere on the instanton geometry, the other for the tunneling to the other side of the instanton geometry.  On the instanton geometry, we use the maximum 2-sphere to separate the two horizon volumes before and after the tunneling.  The downward transition obviously involves a smaller thermal fluctuation.
\label{fig-thermalCDL}}
\includegraphics[width=8cm]{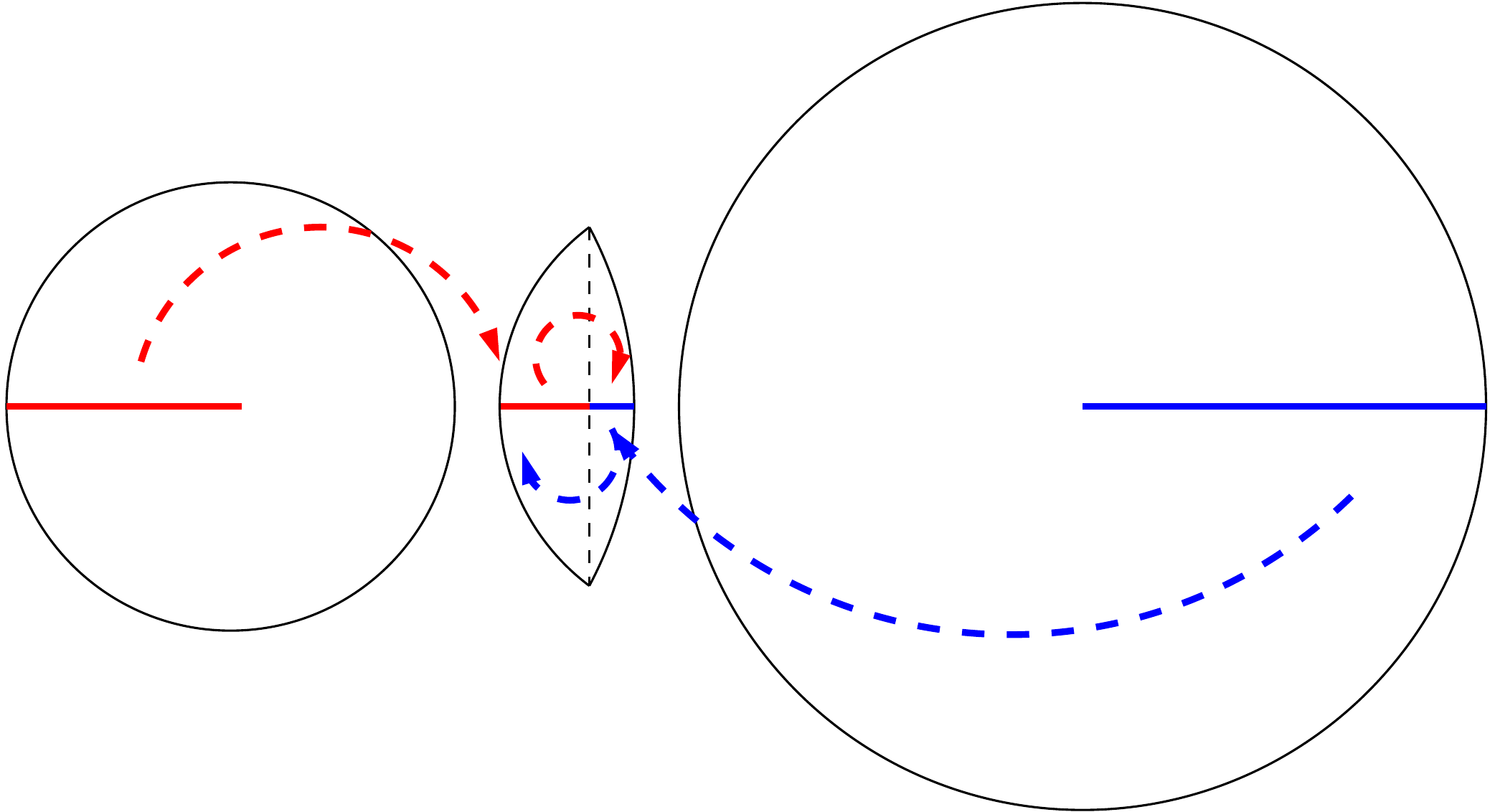}
\caption{The same figure for the type B CDL instanton.  Here, both downward and upward tunnelings involve some dramatic thermal fluctuations to begin with.
\label{fig-thermalVIS}}
\end{center}
\end{figure}

This interpretation clarified a few confusions.  For example, without an exactly thin wall, the ``false vacuum region'' of the instanton will not be identical to the same portion of the false vacuum 4-sphere.  If one takes the instanton as a global geometry, it is unsatisfying that nucleating a bubble requires changes far away, out of causal contact from the bubble.  In the thermal interpretation this has a clear explanation.  With nonzero temperature, the transition is not purely quantum, but always thermally assisted, as depicted in Fig.\ref{fig-thermalCDL}.  The horizon volume of the false vacuum always needs to be thermally excited, even just a little bit, to the configuration that is the left hand side slicing of an instanton, then the quantum tunneling starts.

Note that not only the field configuration of the instanton is slightly away from the pure vacuum, so is the geometry.  This is also straight forward since the gravitational back reaction from a non-vacuum state leads to a non-vacuum geometry\footnote{Therefore, it remains to be clarify as how the equator of a instanton is cut into the initial and final ``horizon volume''.  We will take the mathematically obvious choice: the boundary is the maximum 2-sphere.  The physical reason behind this choice might worth further investigations.}.  Realizing this fact also means that we can accept the reverse tunneling being mediated by the same instanton, just in the reverse direction.  It is a dramatic fluctuation from the true vacuum to the initial condition of this reverse tunneling, in terms of both the field configuration and the geometry.  From the horizon volume of the true vacuum, the fluctuation leads to a bubble of true vacuum surrounded by the false vacuum, and a much reduced horizon size\footnote{We thank Adam Brown for a brief discussion about this.}.  But that is just what is has to be and most of the suppression in the tunneling rate is indeed a thermal factor.

With these in mind, the type B instanton mediates tunneling in the same way, only that both directions require a dramatic thermal fluctuation.  From a horizon volume of a vacuum, we need a thermal fluctuation up to a smaller volume surrounded by a domain wall before the quantum tunneling starts.  This is shown in Fig.\ref{fig-thermalVIS}.

\section{Conclusion}
\label{sec-conclusion}

We explicitly constructed the negative mode for type B CDL instantons.  It is the same radial negative mode as changing the bubble size in the type A instantons.  This natural physical deformation was not considered in earlier literature due to an unnecessary restriction of the geometry.  We removed such restriction by introducing a purely geometric junction in the off-shell configurations of the path integral.  We argued that they satisfy exactly the same principles for gravitational path integral as the original CDL mode analysis.  We also provided simple thick-wall analysis in the Appendix to further justify this novel usage.  Our result agrees with the numerical thick wall examples in\cite{GraTur00,Lav06,BatLav12}.  Although a full analysis including thick-wall effects and less symmetries is still lacking, we believe the conceptual difference between type A and type B instantons is eliminated.  The type A instantons, being similar to the Coleman instantons in flat space, has been widely accepted as the correct saddle point for the tunneling.  The same should be true for the type B instantons.  In the thermal interpretation, we provided the conceptual unification of how both types of instantons mediate upward and downward tunnelings.

\newpage
\acknowledgments
I am especially grateful that Erick Weinberg has planted this problem deeply in my mind, and the discussions we shared to sharpen the argument.  I also thank Adam Brown, Bartek Czech, Ben Freivogel, George Lavrelashvili, Jean-Luc Lehners, Neil Turok, and Xiao Xiao for stimulating discussions.  This work is part of the research program of the Foundation for Fundamental Research on Matter (FOM), which is part of the Netherlands Organization for Scientific Research (NWO).

\bibliographystyle{utcaps}
\bibliography{all}

\providecommand{\href}[2]{#2}\begingroup\raggedright\begin{thebibliography}{10}

\bibitem{CDL}
S.~Coleman and F.~D. Luccia, ``Gravitational effects on and of vacuum decay,''
  {\em Phys. Rev. D} {\bf 21} (1980)  3305--3315.

\bibitem{Col77}
S.~Coleman, ``THE FATE OF THE FALSE VACUUM. 1. {S}EMICLASSICAL THEORY,'' {\em
  Phys. Rev. D} {\bf 15} (1977)  2929--2936.

\bibitem{MarTur07}
K.~Marvel and N.~Turok, ``{Horizons and Tunneling in the Euclidean False
  Vacuum},''
\href{http://arxiv.org/abs/0712.2719}{{\tt arXiv:0712.2719 [hep-th]}}.

\bibitem{Col87}
S.~R. Coleman, ``{QUANTUM TUNNELING AND NEGATIVE EIGENVALUES},''
\href{http://dx.doi.org/10.1016/0550-3213(88)90308-2}{{\em Nucl.Phys.} {\bf
  B298} (1988)  178}.

\bibitem{TanSas92}
T.~Tanaka and M.~Sasaki, ``{False vacuum decay with gravity: Negative mode
  problem},'' {\em Prog.Theor.Phys.} {\bf 88} (1992)  503--528.

\bibitem{Tan99}
T.~Tanaka, ``{The No - negative mode theorem in false vacuum decay with
  gravity},'' \href{http://dx.doi.org/10.1016/S0550-3213(99)00369-7}{{\em
  Nucl.Phys.} {\bf B556} (1999)  373--396},
  \href{http://arxiv.org/abs/gr-qc/9901082}{{\tt gr-qc/9901082}}.

\bibitem{Lav99}
G.~V. Lavrelashvili, ``{Negative mode problem in false vacuum decay with
  gravity},'' {\em Nucl.Phys.Proc.Suppl.} {\bf 88} (2000)  75--82,
  \href{http://arxiv.org/abs/gr-qc/0004025}{{\tt gr-qc/0004025}}.

\bibitem{KhvLav00}
A.~Khvedelidze, G.~V. Lavrelashvili, and T.~Tanaka, ``{On cosmological
  perturbations in closed FRW model with scalar field and false vacuum
  decay},'' \href{http://dx.doi.org/10.1103/PhysRevD.62.083501}{{\em Phys.Rev.}
  {\bf D62} (2000)  083501}, \href{http://arxiv.org/abs/gr-qc/0001041}{{\tt
  gr-qc/0001041}}.

\bibitem{BatLav12}
L.~Battarra, G.~Lavrelashvili, and J.-L. Lehners, ``{Negative Modes of
  Oscillating Instantons},'' \href{http://arxiv.org/abs/1208.2182}{{\tt
  1208.2182}}.

\bibitem{GraTur00}
S.~Gratton and N.~Turok, ``Homogeneous modes of cosmological instantons,'' {\em
  Phys. Rev. D} {\bf 63} (2001)  123514,
\href{http://arxiv.org/abs/hep-th/0008235}{{\tt hep-th/0008235}}.

\bibitem{Lav06}
G.~Lavrelashvili, ``{The Number of negative modes of the oscillating
  bounces},'' \href{http://dx.doi.org/10.1103/PhysRevD.73.083513}{{\em
  Phys.Rev.} {\bf D73} (2006)  083513},
  \href{http://arxiv.org/abs/gr-qc/0602039}{{\tt gr-qc/0602039}}.

\bibitem{BroWei07}
A.~R. Brown and E.~J. Weinberg, ``{Thermal derivation of the Coleman-De Luccia
  tunneling prescription},''
  \href{http://dx.doi.org/10.1103/PhysRevD.76.064003}{{\em Phys. Rev.} {\bf
  D76} (2007)  064003},
\href{http://arxiv.org/abs/0706.1573}{{\tt arXiv:0706.1573 [hep-th]}}.

\bibitem{LeeWei87}
K.~Lee and E.~J. Weinberg, ``Decay of the true vacuum in curved space-time,''
  {\em Phys. Rev. D} {\bf 36} (1987)  1088--1094.

\bibitem{Par83}
S.~Parke, ``Gravity and the decay of the false vacuum,'' {\em Physics Letters
  B} {\bf 121} (1983) no.~5, 313 -- 315.

\bibitem{Isr66}
W.~Israel, ``{Singular hypersurfaces and thin shells in general relativity},''
\href{http://dx.doi.org/10.1007/BF02730328}{{\em Nuovo Cim.} {\bf B44S10}
  (1966)  1}.

\bibitem{HarSor81}
J.~Hartle and R.~Sorkin, ``{BOUNDARY TERMS IN THE ACTION FOR THE REGGE
  CALCULUS},''
\href{http://dx.doi.org/10.1007/BF00757240}{{\em Gen.Rel.Grav.} {\bf 13} (1981)
   541--549}.

\bibitem{Hay93}
G.~Hayward, ``{Gravitational action for space-times with nonsmooth
  boundaries},''
\href{http://dx.doi.org/10.1103/PhysRevD.47.3275}{{\em Phys.Rev.} {\bf D47}
  (1993)  3275--3280}.

\end{thebibliography}\endgroup

\appendix

\section{Thick Wall Constructions}
\label{sec-app}

The idea of a ``purely geometric junction'' appeared much earlier, and the calculation of their contribution to the action is well-known\cite{HarSor81,Hay93}.  However it might be the first time that they play a crucial role in evaluating the off-shell value of Euclidean action.  Certain level of scrutiny is warranted.  Two reasonable concerns were brought to our attention independently and separately by Brown, Freivogel, Weinberg and Xiao.  Here we provide the thick wall justification of our thin wall calculation to address these concerns.

\subsection{Thick-Wall Field Profile}
\label{sec-thick}

A great deal of subtleties in gravitational path integral come from the constraint, Eq.~(\ref{eq-constraint}).  We imagined a purely geometric junction of zero thickness and avoided any explicit consequence from the constraint.  One might worry that we are implicitly violating the constraint thus the configurations studied are not allowed in the path integral.  The specific objection goes like the following.
\ \\ \ \\
{\it The Objection.} \\

Consider a purely geometric junction that connects a shell
\begin{equation}
\rho = R \sin\frac{\xi}{R}~
\label{eq-outside}
\end{equation}
up to some $\bar{\xi}$ with its mirror image.  The radius of this junction is of course smaller than $R$. 
\begin{equation}
\bar{\rho} = R \sin\frac{\bar{\xi}}{R}<R~.
\end{equation}

Now imagine a thick wall version of this, there must be a place that $\rho'=0$ because it changes sign, and this must happen at some value close to $\bar\rho<R$.  Therefore, the constraint equation, Eq.~(\ref{eq-constraint}), demands that at this point, the field could not have stayed in the vacuum.  The idea of a ``purely geometric junction'' is wrong since a nontrivial field profile is necessary.
\ \\ \ \\
{\it The Answer.} \\

It is certainly true that any thick wall geometric junction cannot be pure---certain field profile must accompany it to obey the constraint.  However, all we cared about was to evaluate its contribution to the action.  When the accompanying field profile contributes to a small correction to the value of this purely geometric junction, our method is still valid.  That is indeed the case when the potential allows thin-wall approximation.

\begin{figure}
\begin{center}
\includegraphics[width=8cm]{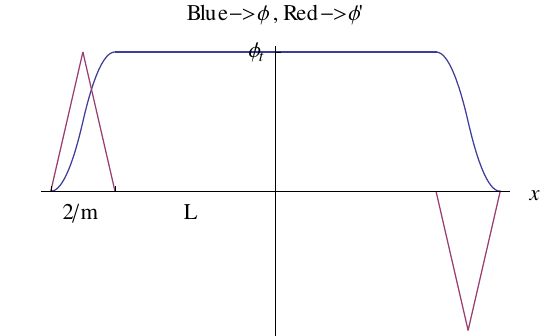}
\caption{The blue curve shows the field profile that stays at $\phi_t$ during the interval of $2L$, with transitions from and to zero within the time scale of $2m^{-1}$.  The red curve is the $\phi'$ profile. 
\label{fig-thickfield}}
\end{center}
\end{figure}

First we expand Eq.~(\ref{eq-pot}), including the uplift to de~Sitter, near one vacuum.
\begin{equation}
V(\phi) = \frac{m^2}{2}\phi^2 + \frac{3M_p^2}{R^2}~,
\label{eq-potA}
\end{equation}
where $m^2 = 8\phi_0^2\lambda^2$ and the definition of $\phi$ is shifted.  Instead of directly matching two shells, we will insert a narrow segment in the middle.  We will replace the coordinate $\xi$ by $x$ within this segment, where $x=0$ sits the middle.  Note that we do not need to obey Eq.~(\ref{eq-field}), so basically we are just inventing a field configuration that solves Eq.~(\ref{eq-constraint}) for our purpose.  The field configuration we want should be continuous in $\phi'$.  The most na\"ive description is from the acceleration, 
\begin{eqnarray}
\phi''(x) &=& 2m^2\phi_t~, \ \ \ \ \ \ {\rm for} \ \ -L-2m^{-1}<x<-L-m^{-1}~, 
\nonumber \\
         &=& -2m^2\phi_t~, \ \ \ {\rm for} \ \ -L-m^{-1}<x<-L~,
\nonumber \\
         &=& 0~, \ \ \ \ \ \ \ \ \ {\rm for} \ \ -L<x<L~,
\nonumber \\
         &=& -2m^2\phi_t~, \ \ \ {\rm for} \ \ L<x<L+m^{-1}~,
\nonumber \\
         &=& 2m^2\phi_t~, \ \ \ \ \ \ {\rm for} \ \ L+m^{-1}<x<L+2m^{-1}~.
\end{eqnarray}
Namely, $\phi$ smoothly increases from $0$ to $\phi_t$ during a short interval $2m^{-1}$, stays at that value for $2L$, then decreases back to zero, as shown in Fig.~\ref{fig-thickfield}.

The value of $\phi_t$ determines the geometry during the middle $2L$ interval.
\begin{eqnarray}
\rho(x) &=& \rho_0 \cos\frac{x}{\rho_0}~, \label{eq-mid} \\
\frac{3M_p^2}{\rho_0^2} &=& \frac{3M_p^2}{R^2}+\frac{m^2}{2}\phi_t^2~.
\label{eq-rho0}
\end{eqnarray}
The purpose of this purely geometric junction is to hold the place of the domain wall in the on-shell configuration.  The required $\rho_0$ is given by
\begin{equation}
\frac{3M_p^2}{\rho_0^2} = V_{\rm top}-\frac{\phi_{ins}'^2}{2}~.
\label{eq-place}
\end{equation}
Here $V_{\rm top}$ is the top of the potential barrier, and $\phi_{ins}'$ is the field velocity there in the instanton solution.  Directly comparing Eq.~(\ref{eq-rho0}) and (\ref{eq-place}), we already see that $\phi_t$ does not need to reach the top of the potential barrier.  Actually, for a potential allowing the thin-wall approximation, $\phi_{ins}'$ is large in the sense that
\begin{equation}
V_{\rm top}-\frac{3M_p^2}{R^2}\gg 
V_{\rm top}-\frac{\phi_{ins}'^2}{2}-\frac{3M_p^2}{R^2}~.
\label{eq-thin}
\end{equation}
Thus the required $\phi_t$ in Eq.~(\ref{eq-rho0}) is far away from the top of the potential barrier and remains in the region that the approximation, Eq.~(\ref{eq-potA}), is valid.

The thin-wall requirement also means $R\gg m^{-1}$.  Combined with the fact that during the $2m^{-1}$ interval when $\phi$ changes, the relevant change in Eq.~(\ref{eq-constraint}) is bounded,
\begin{equation}
\left|\left(\frac{\phi'^2}{2}-V\right)+\frac{3M_p^2}{R^2}\right|
<m^2\phi_0^2=6M_p^2\left(\frac{1}{\rho_0^2}-\frac{1}{R^2}\right)~,
\end{equation}
we know $\rho$ does not change too much during this interval.  This means the geometry of the inserted segment, Eq.~(\ref{eq-mid}), matches to the two shells, Eq.~(\ref{eq-outside}), roughly by
\begin{equation}
\rho_0\cos\frac{L}{\rho_0}=\bar\rho~.
\end{equation}

Now we can calculate the action contribution of this middle segment.
\begin{eqnarray}
S_{\rm mid}&=&\int_{-L-2m^{-1}}^{L+2m^{-1}} \rho^3 dx
\left[\left(\frac{\phi'^2}{2}+V\right) - 3M_p^2\frac{1-\rho'^2-\rho\rho''}{\rho^2} \right] \\
&=& L_{g,wall}(R,R,\bar\rho) + 
\int_{-L-2m^{-1}}^{L+2m^{-1}} \rho^3 dx
\left(2V - \frac{6M_p^2}{\rho^2}\right)~.
\label{eq-thick}
\end{eqnarray}
We have integrated by part to get the boundary term that exactly equals to the contribution from a purely geometric junction.  Now obviously, the middle range $L$ is just a place holder.  We can take $\rho_0\rightarrow\bar\rho$, such that $L\rightarrow0$.  This extra term is an integral similar to other terms in the action, but with a small integration range, $2m^{-1}\ll R$.  Therefore we can see that typically, namely for an order one geometric junction, $\rho_0\sim R$, the extra integral is a small correction to the purely geometric term.

This argument will not apply in two extreme cases.  First when the bubble (which this purely geometric junction is supposed to hold place for) is originally small, $\rho_0\rightarrow m^{-1}$, then $L_{g,wall}$ itself becomes small and the integral term is not negligible.  However this limit means the domain wall thickness is comparable to $\rho_0$, which is exactly when the thin-wall approximation breaks down.  For that case a more complete thick-wall analysis is needed and our approach was never meant to be valid anyway.  The other limit is when the purely geometric junction happens to be very mild, $\rho_0\rightarrow R$ and $L_{g,wall}$ is again close to zero.  In that case the field contribution will not be negligible.  That is actually crucial in the next section to resolve a paradox.  Here we are satisfied that away from these two extremes, an order one purely geometric junction is an appropriate approximation of a thick wall object that satisfies the constraint equation.

\subsection{Resolving an Apparent Paradox}
\label{sec-paradox}

Another way to see potential problems with the purely geometric junction is the following paradox.
\ \\ \ \\
{\it The Paradox.} \\

\begin{figure}
\begin{center}
\includegraphics[width=8cm]{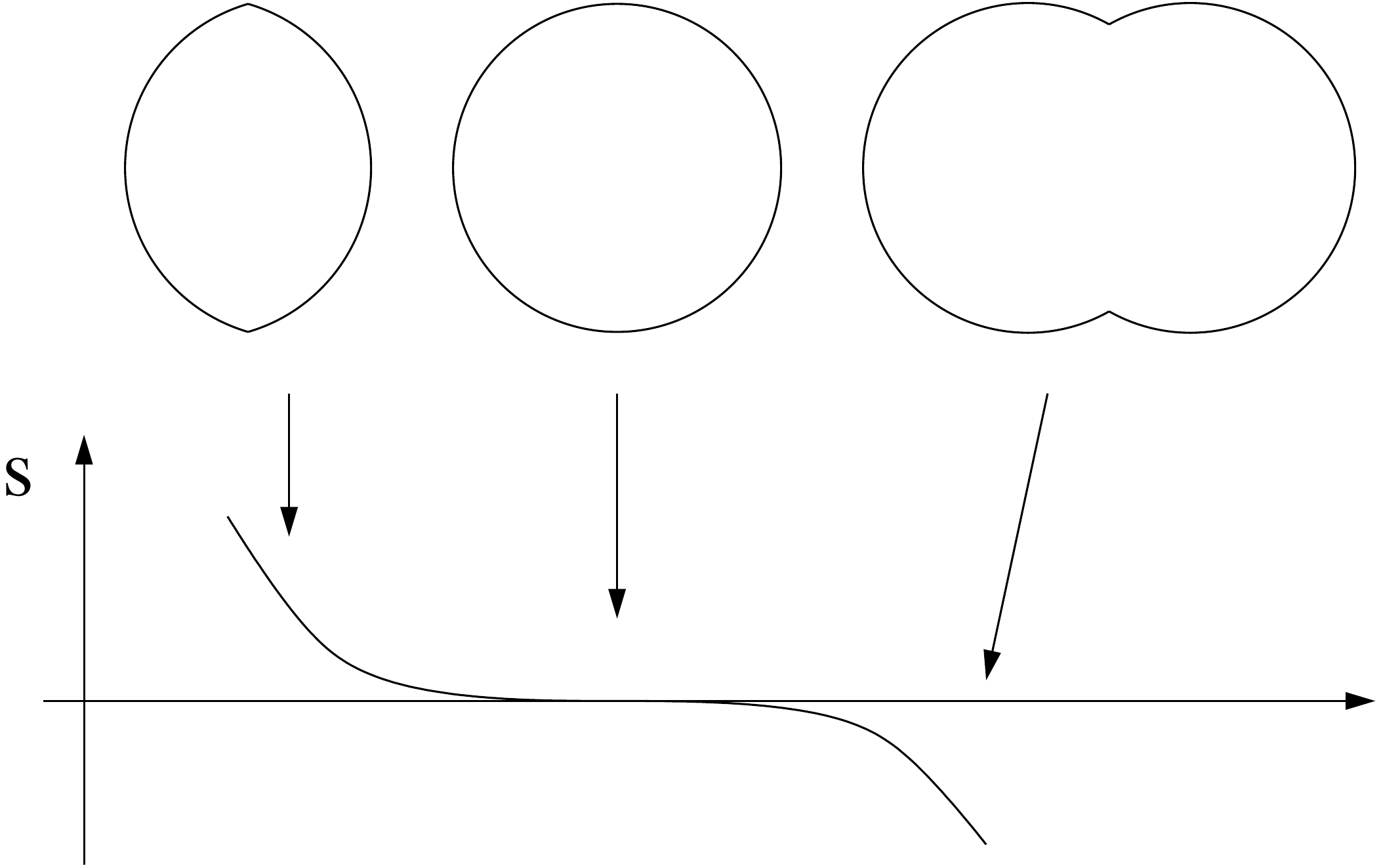}
\caption{Action as a function of the length $l$ of the geometry.  Without the geometric junction it is a 4-sphere, $l=2R$.  A convex junction makes it smaller, and a concave junction makes it larger.  The behavior of the action is $const.+(2R-l)^3$ near $l=2R$.
\label{fig-paradox}}
\end{center}
\end{figure}

Consider the vacuum solution of the potential given by Eq.~(\ref{eq-potA}). It is a 4-sphere with radius $R$.  Now imaging that we develop a purely geometric junction on the equator.  For a convex type of junction that both sides are smaller than half of the 4-sphere, the action will be higher; for the concave type that both sides are bigger than half of the 4-sphere, the action will be lower. (which eventually splits into two 4-spheres)  So it seems like the purely geometric junction introduces a fictitious instability.  It is actually a 3rd order marginal instability as shown in Fig.\ref{fig-paradox}.  

\ \\ \ \\
{\it The Answer.}\\

As hinted in the previous section, the field contribution to a very mild junction cannot be ignored.  If we try to fluctuate these junctions from nothing, we have to keep track of the total action in the thick wall analysis.  We will then see that the field contribution is lower order and positive in the direction of this marginal instability, therefore cures it.

Consider attaching two semi-4-spheres to a middle band, with a waist radius $\rho_w<R$, as the smooth version of a concave geometric junction.  As a small fluctuation, this band should last for an interval $L\ll R$, dips only a little bit $\Delta\rho=(R-\rho_w)\ll R$, and involves a small field fluctuation $m^2\phi_w^2\ll 3M_p^2/R^2$.

The thick wall action contribution of the waist is
\begin{eqnarray}
S_{\rm mid}&=&\int_{-L}^{L} \rho^3 dx
\left[\left(\frac{\phi'^2}{2}+V\right) - 3M_p^2\frac{1-\rho'^2-\rho\rho''}{\rho^2} \right] \\
&=& \int_{-L}^{L} \rho^3 dx
\left(2V - \frac{6M_p^2}{\rho^2}\right)~.
\end{eqnarray}
We have again integrated by part, but this time the boundary term is zero because the waist connects to the equator of two hemispheres where $\rho'=0$.

According to the constraint, Eq.~(\ref{eq-constraint}), we can rewrite the integrand as
\begin{equation}
2V - \frac{6M_p^2}{\rho^2}=\phi'^2-6M_p^2\frac{\rho'^2}{\rho^2}~.
\end{equation}
We can see that the $\phi'^2$ term is positive definite and may be the cure we want.  In order to prove that in general it will, we should make the assumption to minimize it.  Namely, we minimize the number of wiggles in $\phi$ profile such that it monotonically increases to $\phi_w$ in the middle, then monotonically decreases back to zero.  At the matching points to the two shells, and at the middle of the waist, the above quantity is zero since all derivatives are zero.  So we can estimate this integrand by how it ``grows'' from the matching point to the waist. 
\begin{eqnarray}
|\phi'| &\sim& \frac{\phi_w}{L}~, \\
|\rho'| &\sim& \frac{R-\rho_w}{L}=\frac{\Delta\rho}{L}~.
\end{eqnarray}
Plugging the above estimators into the integrand, we get
\begin{equation}
2V - \frac{6M_p^2}{\rho^2} =
\frac{6M_p^2}{R^2L^2}
\left(\frac{2\Delta\rho}{m^2R}-\Delta\rho^2\right)~.
\end{equation}
We can see that in the limit $\Delta\rho\rightarrow0$, this is a positive definite quantity.  So the thick wall contribution cures the apparent marginal instability in the thin-wall analysis.

\end{document}